\documentclass[letterpaper,journal]{IEEEtran}
\usepackage{cite}
\usepackage{amsmath,amssymb,amsfonts}
\usepackage{algorithmic}
\usepackage{algorithm}
\usepackage{graphicx}
\usepackage{textcomp}
\usepackage{makecell}
\usepackage{booktabs}
\usepackage{tabularx}
\usepackage{pifont} 
\newcommand{\cmark}{\ding{51}}
\newcommand{\xmark}{\ding{55}}
\newcommand{\partialmark}{$\triangle$} 
\usepackage{setspace}
\def\BibTeX{{\rm B\kern-.05em{\sc i\kern-.025em b}\kern-.08em
    T\kern-.1667em\lower.7ex\hbox{E}\kern-.125emX}}
\begin{document}

\title{A Decentralized Root Cause Localization Approach for Edge Computing Environments}

\author{Duneesha Fernando, 
Maria A. Rodriguez, 
Rajkumar Buyya
\thanks{D. Fernando, M. A. Rodriguez, and R. Buyya are with the Quantum Cloud Computing and Distributed Systems (qCLOUDS) Laboratory, School of Computing and Information
Systems, The University of Melbourne, Parkville, Australia.}
}

\maketitle

\begin{abstract}
Edge computing environments host increasingly complex microservice-based IoT applications, which are prone to performance anomalies that can propagate across dependent services. Identifying the true source of such anomalies—known as Root Cause Localization (RCL)—is essential for timely mitigation. However, existing RCL approaches are designed for cloud environments and rely on centralized analysis, which increases latency and communication overhead when applied at the edge. This paper proposes a decentralized RCL approach that executes localization directly at the edge device level using the Personalized PageRank (PPR) algorithm. The proposed method first groups microservices into communication- and colocation-aware clusters, thereby confining most anomaly propagation within cluster boundaries. Within each cluster, PPR is executed locally to identify the root cause, significantly reducing localization time. For the rare cases where anomalies propagate across clusters, we introduce an inter-cluster peer-to-peer approximation process, enabling lightweight coordination among clusters with minimal communication overhead. To enhance the accuracy of localization in heterogeneous edge environments, we also propose a novel anomaly scoring mechanism tailored to the diverse anomaly triggers that arise across microservice, device, and network layers. Evaluation results on the publicly available edge dataset, MicroCERCL, demonstrate that the proposed decentralized approach achieves comparable or higher localization accuracy than its centralized counterpart while reducing localization time by up to 34\%. These findings highlight that decentralized graph-based RCL can provide a practical and efficient solution for anomaly diagnosis in resource-constrained edge environments.
\end{abstract}


\section{Introduction}
Edge-cloud integrated environments consist of geographically distributed devices with heterogeneous computing, storage, and networking capabilities. IoT applications are frequently architected as microservices and deployed in these distributed environments to satisfy the Quality of Service (QoS) requirements of each module while optimizing resource utilization \cite{al-doghman2023ai-enabled, WU2023Towards}. In addition to QoS-aware placement, microservices in IoT applications are also commonly deployed based on communication frequency—i.e., microservices that frequently interact are co-located to reduce communication overhead and latency \cite{zhu2024microcercl}. While QoS-aware placement serves as the primary strategy for deploying microservices in heterogeneous edge environments, communication-aware placement generally acts as a complementary secondary approach. 

Over time, these microservice-based IoT applications are susceptible to performance anomalies caused by resource hogging (e.g., CPU or memory) and resource contention, which can negatively impact their QoS and violate their Service Level Agreements \cite{becker2020towards, Soualhia2019infrastructure}. Therefore, it is crucial to conduct performance anomaly detection on microservice-based IoT applications in edge computing environments and eventually mitigate such anomalies. 

However, anomalies can propagate through communication and colocation dependencies in microservice architectures \cite{scheinert2021arvalus, tian2023microgbpm}. In other words, an anomaly originating in one microservice can cascade to others: for example, if a data aggregation microservice running on an edge node suffers from memory saturation, it may delay the delivery of processed sensor data. This slowdown can then cause dependent services, such as real-time anomaly detectors or actuator controllers, to experience increased latency or dropped requests, which in turn may appear anomalous even though they are not the true source. As a result, the system may encounter multiple detections across the architecture, even though only a single source is responsible for triggering the anomaly. Identifying the true source of the anomaly is crucial for implementing effective mitigation strategies. This challenge is known as Root Cause Localization (RCL), and our paper focuses specifically on addressing this problem.

Existing approaches for RCL in microservices primarily focus on cloud environments \cite{soldani2022survey, fu2025intelligentrclsurvey, wang2024comprehensivesurveyrootcause}. At the time of writing, there is only one study on RCL aimed at edge computing environments, known as MicroCERCL \cite{zhu2024microcercl}. Both cloud-based RCL approaches and MicroCERCL rely on telemetry data collected from decentralized monitoring agents and transmitted to a central location for analysis. While effective in cloud settings, applying such a centralized approach at the edge increases data transfer times, thereby degrading localization speed. This highlights an opportunity to develop solutions that execute RCL closer to edge devices. To address this gap, our research proposes a decentralized RCL approach that performs the analysis directly at the edge device level.

Graph-based methods represent the current state of the art in RCL \cite{soldani2022survey, fu2025intelligentrclsurvey, wang2024comprehensivesurveyrootcause}. Within this space, unsupervised heuristic approaches that exploit graph centralization play a particularly important role. In an RCL setting, centralization measures help identify the most “influential” or “central” node in the anomaly propagation graph, under the assumption that the true root cause will often lie at or near the center of anomalous interactions. These approaches are lightweight and offer high interpretability, making them well-suited for resource-constrained edge environments when compared to more computationally intensive supervised deep learning methods, such as Graph Neural Networks (GNNs) \cite{zhu2024microcercl}. Among the unsupervised centralization techniques, the Personalized PageRank (PPR) algorithm has proven to be a particularly effective method for RCL \cite{wu2020microrca, tian2023microgbpm}. Building upon this foundation, we employ PPR as the core algorithm in our work.

Edge environments, however, are characterized by a large number of highly distributed devices, creating a complex problem space that graph-based approaches must navigate. This complexity can significantly increase localization times \cite{bulla2019edgercl, kalinagac2023liability, scheinert2021arvalus}. To address this challenge, our proposed PPR-based decentralized RCL approach groups microservices into clusters based on their communication and colocation dependencies. This means that when an anomaly occurs, it typically propagates within the identified cluster boundaries. In such cases, the source microservice can be located by executing PPR only within that cluster. This reduces the search space that the PPR algorithm needs to explore, leading to shorter localization times compared to the traditional centralized approaches. 

In rare instances where anomalies may propagate outside of the cluster, we propose an inter-cluster peer-to-peer approximation process. In our inter-cluster peer-to-peer approximation process, clusters exchange the average of their anomaly scores, which serves as a compact representation of the anomaly state of each cluster. This exchanged value is incorporated into the receiving cluster’s anomaly score list as an approximate indicator of the other cluster’s influence. Each cluster then executes its PPR-based RCL algorithm independently. By doing so, we limit communication overhead while still retaining the ability to capture inter-cluster anomaly propagation across iterations. 

Existing PPR-based RCL approaches designed for cloud environments typically compute anomaly scores by quantifying their correlation with anomalous response time metrics \cite{wu2020microrca, zhang2021aamr}. While this works well in cloud settings, such correlation-based scoring is less effective in IoT and edge environments, where anomalies can originate from heterogeneous layers of the infrastructure. As a secondary contribution, we extend the conventional PPR framework by introducing a novel anomaly scoring mechanism tailored to the characteristics of edge environments. To the best of our knowledge, this is the first work to propose an RCL approach tailored to the characteristics and limitations of edge environments while specifically trying to reduce the impact of scalability on localization time through decentralization.

We evaluated our proposed decentralized RCL approach using a publicly available dataset released alongside the MicroCERCL paper \cite{zhu2024microcercl}. Evaluations performed on the dataset using PPR with our proposed anomaly scoring mechanism demonstrate that the decentralized RCL approach achieves the same level of accuracy as its centralized counterpart, while significantly reducing localization time in scenarios of single-cluster anomaly propagation, as well as in rare instances where anomalies propagate across multiple clusters.  

The rest of the paper is organized as follows: Section \ref{sec:lit_review} reviews current research on RCL in microservice-based systems, discusses their deployment models, presents an overview of the Personalized PageRank algorithm, and examines the efficiency aspects of existing RCL techniques. Section \ref{sec:methodology} introduces our proposed decentralized RCL methodology, including the novel anomaly scoring mechanism, the communication- and colocation-based clustering approach, and the decentralized execution of the PPR algorithm. Section \ref{sec:perf_eval} presents the performance evaluation, comparing our approach against the centralized baseline and analyzing both single-cluster and multi-cluster anomaly propagation scenarios. Section \ref{sec:conclusions} concludes the paper and outlines directions for future work.

\section{Background and Related Work}
\label{sec:lit_review}

In this section, we provide an overview of existing RCL techniques, examine their deployment patterns and the challenges of adapting cloud-based approaches to edge environments, present the fundamentals of the PPR algorithm, and discuss the efficiency aspects of existing RCL techniques.
\subsection{Root cause localization techniques}

RCL in microservices environments leverages various data sources to identify performance issues, categorized into metric-based, trace-based, log-based, and multi-source approaches \cite{fu2025intelligentrclsurvey, soldani2022survey}. Metric-based RCL focuses on key performance indicators like CPU utilization and Queries Per Second (QPS), providing direct insights into potentially faulty modules. It stands out for its efficiency and adaptability compared to log-based approaches, which are often voluminous, unstructured, heterogeneous in format, and complex to analyze in real-time \cite{wang2024comprehensivesurveyrootcause, xin2023causalrca}. 

Trace-based RCL, while useful for understanding request flows, is often too coarse-grained to effectively diagnose deeper system issues, such as resource bottlenecks or internal microservice misbehaviors \cite{zhu2024hemirca}. Additionally, these methods encounter scalability challenges, as they require expensive and time-consuming data collection and processing. Furthermore, they require instrumentation of source code, whereas metrics can be collected without intrusive modifications. 

Emerging multi-source RCL techniques aim for a comprehensive view by integrating metrics, logs, and traces, but add complexity due to differences in data formats, granularity, and temporal alignment \cite{wang2024comprehensivesurveyrootcause, yu2023nezha}. In some cases, one modality may even dominate or obscure the others. In contrast, metric-based approaches offer a structured and simpler framework, making them practical and effective for RCL investigations in dynamic edge-cloud environments.

RCL techniques that rely on metrics as the primary source of data can be broadly divided into two categories: correlation-driven statistical approaches and graph-based approaches \cite{fu2025intelligentrclsurvey, wang2024comprehensivesurveyrootcause}. Correlation-driven statistical approaches are typically triggered when an anomaly detection module identifies unusual response times in the frontend microservice. These methods then compute correlations between the anomalous response times and metrics from other microservices, ultimately reporting the metric with the highest correlation and its corresponding microservice \cite{liu2021microhecl, lin2018microscope}. While these methods are suitable for deployment in cloud environments, where web applications are primarily hosted, they fall short when applied at the edge, where anomalies can arise from any metric across various levels of the infrastructure. Moreover, these methods generally struggle with accuracy due to their reliance on simple pairwise correlations, which may overlook multi-service dependencies and indirect causal paths.

In contrast, graph-based RCL methods, which represent the current state of the art, explicitly model service dependencies and anomaly propagation using graph structures constructed after anomaly detection \cite{wu2020microrca, tian2023microgbpm, zhu2024microcercl, zhang2021aamr}. Two main types of graphs are commonly used: causal graphs, which represent direct cause–effect relationships between metrics, and topological graphs, which capture service interactions based on deployment and runtime traces \cite{fu2025intelligentrclsurvey, soldani2022survey}. This work focuses on topological graphs, as they can be easily constructed using deployment information and traces, making them practical for modeling microservice dependencies, unlike causal graphs that are difficult to construct accurately in dynamic edge-cloud environments. 

Once constructed, these graphs can be analyzed using either unsupervised heuristic methods (e.g., centrality-based approaches) or supervised deep learning methods (e.g., Graph Neural Networks, GNNs) \cite{soldani2022survey, wang2024comprehensivesurveyrootcause}. Centrality-based methods exploit graph properties to identify the most “influential” or “central” node within the anomaly propagation graph, under the assumption that the true root cause often lies near the center of anomalous interactions. These methods are lightweight, interpretable, and require no labeled training data or pre-trained models, making them highly suitable for resource-constrained and heterogeneous edge environments \cite{fu2025intelligentrclsurvey}. In contrast, supervised methods such as GNNs are often computationally expensive, demand large labeled datasets, and require retraining to adapt to new system configurations, which limits their practicality at the edge. Among other unsupervised centralization techniques such as degree centrality (identifies nodes with the largest number of anomalous connections), betweenness centrality (highlights nodes acting as bridges in anomalous paths), and eigenvector-based methods (emphasize nodes connected to other highly anomalous nodes), Personalized PageRank (PPR) has emerged as a particularly effective method for RCL \cite{wu2020microrca,tian2023microgbpm}. Unlike simple degree or betweenness measures, PPR captures both local and global structural dependencies, enabling it to rank potential root causes more robustly even in large and heterogeneous graphs. Therefore, we adopt the PPR algorithm as the foundation for our RCL solution. 

In the next section, we discuss how existing RCL solutions are typically deployed and highlight the challenges that arise when these cloud-based approaches are applied in edge computing environments.

\subsection{Centralized deployment of existing root cause localization solutions}

\begin{figure}[t]
\centerline{\includegraphics[width=\columnwidth]{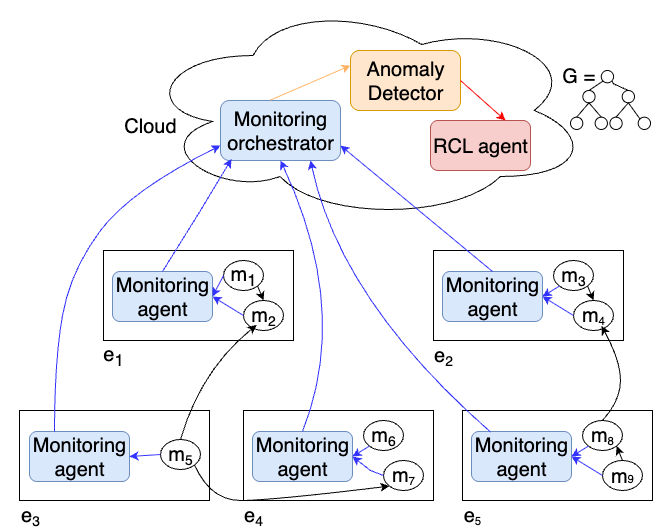}}
\caption{Centralized root cause localization in edge environments}
\label{centralised_rcl}
\end{figure}

As shown in Figure \ref{centralised_rcl}, the edge infrastructure consists of a set of edge devices \[
\mathcal{E} = \{ e_1, e_2, \dots, e_{|\mathcal{E}|} \}.
\]

Let the set of microservices be:
\[
\mathcal{M} = \{ m_1, m_2, \dots, m_{|\mathcal{M}|} \}.
\]

Each microservice \( m \in \mathcal{M} \) is deployed on exactly one edge device. The deployment mapping can be defined as
\[
\delta: \mathcal{M} \to \mathcal{E},
\]
where \(\delta(m)\) gives the device on which microservice \(m\) is deployed.

Each edge device \( e \in \mathcal{E} \) has a monitoring agent that gathers performance and resource consumption metrics from the microservices deployed on that device. Both cloud-based RCL studies and MicroCERCL send the collected telemetry data to a central location, such as the cloud, for analysis. While this approach works well in the cloud, where data transfer times are relatively low, applying it at the edge increases data transfer times, which negatively impacts the speed of localization. Furthermore, transferring data to a central location does not align well with the network instability characteristic of cloud-edge environments \cite{zhu2024microcercl}. Therefore, it is essential to develop solutions that enable RCL to be performed closer to the edge devices in edge computing environments.

In situations where graph-based RCL approaches are in place, the centralized RCL module maintains a topology graph
\[
\mathcal{G} = (\mathcal{V}, \mathcal{L}),
\]
which captures the structure and interactions of the monitored system.

The set of vertices \(V\) consists of both microservices and the edge devices on which they are deployed:
\[
\mathcal{V} = \mathcal{M} \cup \mathcal{E}
\]

The set of edges \(L\) represents relationships between these vertices and is composed of two distinct types:
\[
\mathcal{L} = \mathcal{L}_{\mathrm{comm}} \cup \mathcal{L}_{\mathrm{dep}}
\]

\begin{enumerate}
    \item Communication edges (\(L_{comm}\)) capture interactions between microservices, derived from trace data:
    \[
    \mathcal{L}_{\mathrm{comm}} \subseteq \mathcal{M} \times \mathcal{M},
    \]
    where \((m_i, m_j) \in \mathcal{L}_{\mathrm{comm}}\) if microservice \(m_i\) communicates with \(m_j\).
    \item Deployment edges (\(L_{dep}\)) link each microservice to the edge device on which it is hosted, obtained from the deployment configuration:
    \[
    \mathcal{L}_{\mathrm{dep}} = \{ (m, \delta(m)) \mid m \in \mathcal{M} \}.
    \]
\end{enumerate}

Together, these vertices and edges form a topology that reflects both the logical communication between microservices and their physical placement across edge devices, enabling the RCL process to reason over the system’s structure. 

\begin{figure}[t]
\centerline{\includegraphics[width=\columnwidth]{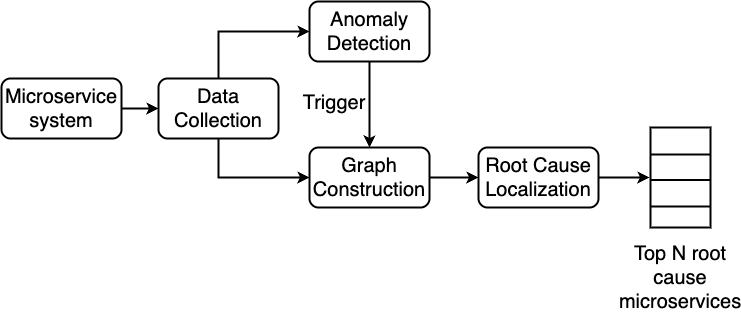}}
\caption{Overview of the root cause localization process}
\label{rcl_workflow}
\end{figure}

In cloud environments, where primarily web applications are deployed, continuous anomaly detection is typically performed only on the response times of user-facing microservices \cite{wu2020microrca, tian2023microgbpm, zhang2021aamr,wu2021microdiag}. When an anomaly is detected in any of those metrics, it triggers the RCL process. In contrast, many edge applications do not interact directly with end users and often follow architectures such as publish–subscribe or streaming. Therefore, instead of relying solely on user-facing response times, anomaly detection must be performed at multiple levels across the infrastructure—including microservice, device, and network layers—to ensure effective root cause localization in edge environments. In this context, anomalies detected at any level should trigger the RCL process. Once an anomaly is identified, the edges and vertices of the topology graph ($\mathcal{G}$) are populated with the anomalous metrics from the detection window. This graph serves as input for the PPR algorithm. Figure \ref{rcl_workflow} demonstrates an overview of the RCL process from data collection to anomalous services ranking. The next subsection explains the PPR algorithm in detail.

\subsection{Personalized PageRank algorithm for root cause localization}
\label{subsec:ppr}

Although the topology graph ($\mathcal{G}$) populated with anomalous metrics serves as the input for the RCL process, the Personalized PageRank (PPR) algorithm specifically operates on two key inputs: the personalization vector ($\mathcal{S}$) and the transition probability matrix ($\mathcal{P}$) \cite{Haveliwala2002ppr}. 

The personalization vector \( S \in \mathbb{R}^{|V|} \) represents the prior probability distribution over the nodes, encoding our belief regarding their likelihood of being the root cause. In the proposed framework, \( S \) is derived from anomaly scores assigned to nodes and edges in the topology graph. Formally, for a node \( v_i \in V \):
\[
S_i = \frac{a(v_i)}{\sum_{v_j \in V} a(v_j)}
\]
where \( a(v_i) \) denotes the anomaly score of node \( v_i \). This ensures that \( S \) is a valid probability distribution:
\[
\sum_{i=1}^{|V|} S_i = 1.
\]

The transition probability matrix \( P \in \mathbb{R}^{|V| \times |V|} \) captures the normalized probabilities of transitioning between nodes based on edge weights. For an edge \( l_{ij} \in L \) connecting nodes \( v_i \) and \( v_j \), its weight is determined by its anomaly score \( a(l_{ij}) \). The transition probability is defined as:
\[
P_{ij} = \frac{a(l_{ij})}{\sum_{v_k \in \mathcal{N}(v_i)} a(l_{ik})},
\]
where \( \mathcal{N}(v_i) \) denotes the set of neighbors of \( v_i \). Consequently, \( P \) is a row-stochastic matrix, ensuring:
\[
\sum_{j=1}^{|V|} P_{ij} = 1 \quad \forall i.
\]

PPR-based RCL approaches designed for cloud environments typically compute the anomaly scores for the nodes and edges of the constructed topology graph—required for calculating \( P \) and \( S \)—by quantifying their correlation with anomalous response time metrics \cite{wu2020microrca, zhang2021aamr}. Specifically, for each anomalous node or edge, the correlation between its observed metric values and the anomalous response time metric is measured, thereby assigning higher scores to components exhibiting stronger associations with the observed performance degradation. However, IoT applications would benefit from a novel anomaly scoring mechanism tailored to address the triggers that arise from different levels of edge infrastructure.

With \( P \) and \( S \) defined, the PPR algorithm iteratively refines a ranking vector \( r \in \mathbb{R}^{|V|} \), representing the steady-state probabilities of each node being the root cause. The update rule is given by:
\[
r^{(k+1)} = \alpha P r^{(k)} + (1 - \alpha) S,
\]
where:
\begin{itemize}
    \item \( \alpha \in (0,1) \) is the damping factor, controlling the trade-off between following graph transitions (\( P r^{(k)} \)) and teleporting to the personalization distribution (\( S \));
    \item \( r^{(k)} \) denotes the ranking vector at iteration \( k \).
\end{itemize}

The algorithm converges when the difference between successive iterations falls below a predefined threshold:
\[
\| r^{(k+1)} - r^{(k)} \| < \varepsilon,
\]
where \( \varepsilon \) denotes the desired precision. Convergence is achieved using \textit{power iteration}, with the number of iterations influenced by both \( \alpha \) and \( \varepsilon \). Lower values of \( \alpha \) typically result in faster convergence due to increased reliance on the personalization vector. 

Given that our approach seeks to improve localization efficiency by reducing localization time, we next review the efficiency aspects of existing RCL techniques.

\subsection{Efficient root cause localization techniques}

Training time, localization time, and resource overhead are the most important efficiency metrics considered in studies on cloud RCL \cite{wang2024comprehensivesurveyrootcause}. These metrics hold equal or even greater significance at the edge. By utilizing the PPR algorithm, we have eliminated the need for training and reduced resource overhead due to its lightweight nature. However, our proposed decentralized RCL approach primarily targets the reduction of localization time.

Localization time can be discussed in relation to the time complexity of the PPR algorithm. The time complexity of the PPR algorithm for a given topology graph \(\mathcal{G} = (\mathcal{V}, \mathcal{L})\) depends on the size of the graph, which can be expressed in terms of its number of edges \cite{Wang2022edgeppr}. At each iteration, the algorithm performs a sparse matrix–vector multiplication, whose computational cost is proportional to the number of edges \(|\mathcal{L}|\). Therefore, the per-iteration complexity is \(\mathcal{O}(|\mathcal{L}|)\). If \(k\) denotes the number of iterations required to reach convergence, the total time complexity becomes \(\mathcal{O}(k \cdot |\mathcal{L}|)\). Thus, the overall convergence time of the PPR algorithm scales linearly with both the number of edges and the number of iterations needed to achieve convergence.

In contrast to conventional PageRank algorithm, which distributes teleportation probabilities equally across all nodes, the PPR approach biases the algorithm towards nodes and edges with higher anomaly scores \cite{Haveliwala2002ppr}. This personalization effectively prioritizes microservices and interactions that are more likely to be the root cause at the outset, thereby reducing the number of iterations required for convergence and improving localization efficiency. Edge environments, however, consist of a large number of devices and are highly distributed, creating a complex problem space for graph-based approaches to navigate \cite{bulla2019edgercl, kalinagac2023liability, scheinert2021arvalus}. In such environments, where the underlying dependency graph can become extremely large, reducing the effective search space—and hence the number of edges considered during computation—is crucial for minimizing localization time. This is where our proposed decentralized RCL approach contributes, by narrowing the search space through cluster-based graph partitioning, which indirectly reduces the number of edges involved in PPR computation and thus accelerates convergence. 

\begin{table*}[t]
\centering
\caption{Summary of cloud and edge RCL techniques: accuracy vs efficiency focus}
\label{tab:rcl_summary}
\small
\begin{tabularx}{\textwidth}{l c c >{\centering\arraybackslash}c >{\centering\arraybackslash}c X}
\hline
\textbf{Work} & \textbf{\makecell{Cloud-only/ \\ Cloud-Edge}} & \textbf{\makecell{Centralized/ \\ Decentralized}} & \textbf{\makecell{Accuracy \\Focus}} & \textbf{\makecell{Efficiency \\Focus}} & \textbf{Remarks} \\
\hline
\cite{kalinagac2023liability, fu2025intelligentrclsurvey, zhu2024microirc} & Cloud-only & Centralized & \cmark & \xmark & Focused solely on accuracy \\
\cite{zhang2021aamr,bulla2019edgercl,ren2023grace} & Cloud-only & Centralized & \cmark & \partialmark & Claim faster inference; efficiency not a design priority \\
MicroHECL \cite{liu2021microhecl} & Cloud-only & Centralized & \partialmark & \cmark & Efficient traversal \& pruning; may compromise accuracy \\
PDiagnose \cite{hou2021pdiagnose} & Cloud-only & Centralized & \partialmark & \cmark & Vote-based localization; improves efficiency; may compromise accuracy \\
MicroCERCL \cite{zhu2024microcercl} & Cloud-Edge & Centralized & \cmark & \xmark & Accuracy focus; long inference time reported\\
Our work & Cloud-Edge & Decentralized & \cmark & \cmark & Graph-based; balances accuracy and efficiency \\
\hline
\end{tabularx}
\vspace{1mm}
\vspace{1mm}
{\raggedright
\footnotesize
\cmark = Primary focus / addressed, \partialmark = Partially addressed / may compromise, \xmark = Not addressed / ignored
\par}
\end{table*}

A comparison of cloud and edge RCL techniques in terms of their focus on efficiency (and accuracy as well) is shown in Table~\ref{tab:rcl_summary}. The majority of studies on cloud RCL have primarily focused on improvements in accuracy, evaluating this single aspect \cite{kalinagac2023liability, fu2025intelligentrclsurvey, zhu2024microirc}. Some research has also addressed efficiency, particularly the localization times of their proposed methods. For instance, AAMR \cite{zhang2021aamr}, MicroEGRCL \cite{bulla2019edgercl}, and Grace \cite{ren2023grace} claim to provide faster inference times. However, none of these studies have specifically designed their approaches with efficiency as a priority. 
In addition, the only study focused on edge RCL, MicroCERCL, did not incorporate efficiency considerations into its approach. Consequently, during efficiency evaluations, it was found that its inference time was longer than that of unsupervised heuristic approaches, largely due to the complexity of the network \cite{zhu2024microcercl}.  

MicroHECL \cite{liu2021microhecl} and PDiagnose \cite{hou2021pdiagnose} are cloud RCL techniques that specifically aim to improve efficiency, particularly by providing faster localization speeds. Both approaches eliminate the need for training, similar to our chosen method. They are centralized approaches; MicroHECL achieves efficiency by efficiently traversing the service dependency graph and using pruning techniques to eliminate irrelevant service calls during anomaly propagation chain analysis, which further enhances efficiency. 
On the other hand, PDiagnose tries to reach efficiency by removing the computationally heavy dependency graph-building phase and utilizing a vote-based localization process 
on an anomaly queue. 
However, both approaches adopt relatively simplistic strategies that may compromise localization accuracy. Given the need to strike an appropriate balance between effectiveness and efficiency, our proposed decentralized method leverages a graph-based approach to enhance accuracy while reducing localization time. Additionally, cloud efficiency techniques like pruning, as used in MicroHECL, complement our proposed approach. 

Building on the insights from existing RCL techniques and their efficiency considerations, the next section details our proposed decentralized RCL methodology, which aims to reduce localization time while maintaining high accuracy.

\section{Methodology}
\label{sec:methodology}

\begin{figure}[t]
\centerline{\includegraphics[width=\columnwidth]{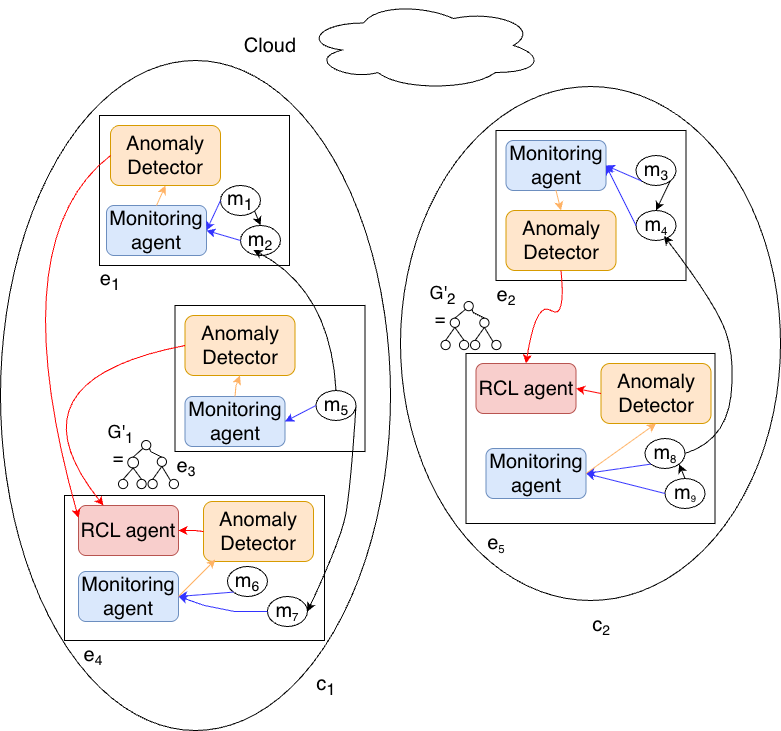}}
\caption{Decentralized root cause localization in edge environments}
\label{decentralised_rcl}
\end{figure}

Our proposed decentralized RCL approach aims to minimize the need for the PPR algorithm to traverse the entire graph by clustering edge devices that frequently communicate with each other. This means that the PPR algorithm only needs to explore the cluster where the anomaly has propagated. As a result, the search space is reduced, leading to shorter localization times. Additionally, since our approach conducts localization as close to the edge device level as possible, it further decreases localization time by minimizing data transfer delays.

We define the set of clusters as 
\[
\mathcal{C} = \{ c_1, c_2, \dots, c_{|\mathcal{C}|} \}.
\]

Each edge device \( e \in \mathcal{E} \) is assigned to exactly one cluster, and we can represent the mapping of edge devices to clusters as
\[
\gamma: \mathcal{E} \to \mathcal{C},
\]
where \(\gamma(e)\) indicates the cluster to which edge device \(e\) belongs. We will provide further details on the clustering algorithm in subsection \ref{subsec:clustering_algo}. 

As illustrated in Figure \ref{decentralised_rcl}, our approach deploys an anomaly detection module at each edge device. This module utilizes a BIRCH clustering-based anomaly detection model. BIRCH is an unsupervised, lightweight technique that is widely used for anomaly detection in multiple root cause localization studies \cite{wu2020microrca, wu2021microdiag, zhang2021aamr, zhu2024microcercl}, and it also suits the properties of edge environments. It analyzes the performance and resource consumption metrics collected by the monitoring agent assigned to each edge device. Unlike cloud-based applications that primarily deploy web applications, anomaly detection for IoT applications—which are not necessarily user-facing—should analyse all metrics obtained from the monitoring agent. 

In our context, where anomaly detection occurs at multiple levels throughout the infrastructure, we consider any detected anomalies as triggers for root cause localization. In the proposed approach, the RCL module corresponding to each cluster is placed at the node with the highest computational power within the cluster. Each RCL module maintains a topology subgraph \( G'_i \) that corresponds to its cluster \( c_i \). When an anomaly is detected, the edges and vertices of the topology subgraph \( G'_i \) are populated with the anomalous metrics gathered from the microservices deployed in that cluster.

Subsequently, we form the personalization vector \(S_i\) and the transition probability matrix \(P_i\) corresponding to the topology subgraph by using a novel anomaly scoring mechanism, which we introduce in subsection \ref{subsec:anomaly_scoring_edge}. The novel anomaly scoring mechanism is tailored to address triggers arising from different levels of the edge infrastructure. Following this, we can execute the PPR algorithm within the cluster. Our approach assumes that in the majority of scenarios, an anomaly will propagate within a single cluster, which means we only need to explore that specific cluster. However, in rare cases where anomalies may propagate outside of the cluster, we provide an inter-cluster peer-to-peer approximation process that the clusters can follow to localize the root cause microservice. This strategy also aims to reduce localization time by parallelizing the graph traversal across clusters and reducing inter-process communication overhead through one time exchange of approximate anomaly scores. The inter-cluster peer-to-peer approximation process will be further explained in subsection \ref{subsec:decentralised_ppr}.

The upcoming sections are organized as follows: Section \ref{subsec:anomaly_scoring_edge} introduces the novel anomaly scoring mechanism tailored for edge environments while section \ref{subsec:clustering_algo} explains the proposed communication and colocation-based clustering approach, followed by the topology subgraph formation algorithm. Section \ref{subsec:decentralised_ppr} explains the decentralized execution of the PPR algorithm for both single cluster and multiple cluster anomaly propagation scenarios.

\subsection{Novel anomaly scoring mechanism for edge environments}
\label{subsec:anomaly_scoring_edge}

PPR-based RCL approaches designed for cloud environments typically compute the anomaly scores for the nodes and edges of the constructed topology graph—required for calculating \( P \) and \( S \)—by quantifying their correlation with anomalous response time metrics \cite{wu2020microrca, zhang2021aamr}. Specifically, for each anomalous node or edge, the correlation between its observed metric values and the anomalous response time metric is measured, thereby assigning higher scores to components exhibiting stronger associations with the observed performance degradation. However, in IoT applications, where anomalies may arise from various levels of edge infrastructure, we introduce a novel anomaly scoring mechanism designed specifically to meet these requirements.

\subsubsection{Anomaly score for microservices}
Each microservice \( m \) is assigned an anomaly score that aggregates two main influences:
\begin{itemize}
\item Anomalous metric influence
\item Incoming anomalous edge influence
\end{itemize}

Thus, the anomaly score for a microservice \( m \), denoted by \( \text{AS}(m) \), is computed as:
\[
\text{AS}(m) = \underbrace{\text{AS}_\text{metric}(m)}_\text{Metric-induced} + \underbrace{\text{AS}_\text{edge}(m)}_\text{Edge-induced}
\]

\paragraph{Anomalous metric influence}
Let \( A \) denote a BIRCH cluster of anomalous time series metrics associated with microservice \( m \). We define the anomalous metric influence score of microservice \( m \), denoted by \( \text{AS}_\text{metric}(m) \), as:

\begin{equation}
\label{eq:as_as_metric}
\text{AS}_\text{metric}(m) = \sum \textit{corr}(A)
\end{equation}

where \( \textit{corr}(A) \) is the pairwise correlation among all metric time series in cluster \( A \) computed using a direction-invariant correlation metric such as the absolute Pearson correlation metric.

\paragraph{Incoming anomalous edge influence}
To capture the performance impact on downstream services due to microservice \( m \), we consider all anomalous incoming edges represented as microservice pairs \( (m_d,m) \), where \( m_d \in D \) is the downstream microservice sending requests to the target microservice \( m \). \( D \) is the set of all downstream microservices from \( m \) with an anomalous incoming edge to \( m \). For each such edge, we compute the correlation between the edge’s anomalous response time (represented by the 90th percentile of latency) and each anomalous metric \( x_i \in A \) of the target microservice \( m \) and obtain the maximum of these correlation values. The sum of all such maximum correlation values is assigned to the incoming anomalous edge influence score of microservice \( m \), denoted by \( \text{AS}_\text{edge}(m) \), as shown in equation \ref{eq:as_edge}.

\begin{align}
\label{eq:as_edge}
\scalebox{1}{
$\text{AS}_\text{edge}(m) = \sum_{m_d \in D} \max_{x_i \in A} \mathit{corr}(\mathrm{RT}_{m_d \to m}, x_i)$
}
\end{align}

This formulation enables the anomaly scoring mechanism to account for both internal anomalies and anomalous behavior observed at communication boundaries, capturing the cascading effect of faults within microservice architectures. These microservice-level scores are used to form the personalization vector \(S\).

\subsubsection{Anomaly score for edges}

The anomaly score for each inter-service edge represented as a microservice pair \((m_s,m_t)\), denoted by \( \text{AS}(m_s,m_t) \), is computed using the same formulation as the anomalous metric influence score presented in equation \ref{eq:as_as_metric}.

\begin{equation}
\text{AS}(m_s,m_t) = \sum \textit{corr}(B)
\end{equation}

\( B \) denotes a BIRCH cluster of anomalous edge-level metrics (e.g., different percentiles of response time) associated with the communication from microservice \(m_s\) to \(m_t\), while \( \textit{corr}(B) \) is the pairwise correlation among all metric time series in cluster \( B \) computed using a direction-invariant correlation metric such as the absolute Pearson correlation metric. 

These edge-level scores are subsequently used to populate the transition probabilities in the matrix \(P\), guiding the flow of anomaly information through the system topology during the RCL phase.

This novel anomaly scoring mechanism, specifically designed for edge computing environments, is utilized in our proposed decentralized RCL approach, which is explained in the upcoming sections.

\subsection{Communication and colocation-based clustering}
\label{subsec:clustering_algo}

Anomalies can propagate through communication and colocation dependencies in microservice architectures \cite{scheinert2021arvalus, tian2023microgbpm}. Our proposed clustering approach aims to group microservices that are both colocated and frequently communicate with one another, so that when an anomaly occurs, it propagates within the identified cluster boundary in most cases.

\begin{algorithm}[t]
\setstretch{0.9}
    \caption{Communication and colocation-based clustering algorithm} 
    \label{algo:comm_col_cluster_algo}
    \begin{algorithmic}[1]
        \STATE{\textit{Input} : Centralized topology graph $G' = (V,L_{comm})$ where $V$ is the set of microservices $M$ and edge devices $E$, and $L_{comm}$ is the set of communication edges between microservices}
        \STATE{\textit{Input} : Function $\delta(m)$ which maps each microservice $m$ to its deployment node}
        \STATE{\textit{Input} : Threshold to merge clusters $\tau$}
        \STATE{\textit{Output} : Cluster to edge device mapping $cluster\_map$}
        \STATE{Initialize $cluster\_map$ such that for each deployment node $e_i$, there is a cluster $c_{e_i} = \{ m \in V \mid \delta(m) = e_i \}$}
        \STATE\label{comm_col_cluster:weight_calc}{For each pair of clusters $(c_{e_i}, c_{e_j})$, define the inter-cluster communication weight $w(c_{e_i}, c_{e_j}) = \sum\limits_{\substack{m_p \in c_{e_i} \\ m_q \in c_{e_j}}} freq(m_p, m_q)$, where $freq(m_p, m_q)$ is the communication frequency between microservices $m_p$ and $m_q$}
        \REPEAT
            \STATE{Find the cluster pair $(c_p, c_q)$ with maximum weight $w(c_p, c_q)$}
            \IF{$w(c_p, c_q) > \tau$}
                \STATE{Merge $c_p$ and $c_q$ into a new cluster $c_{new} = c_p \cup c_q$}
                \STATE{Update $cluster\_map = (cluster\_map \setminus \{c_p, c_q\}) \cup \{c_{new}\}$}
                \STATE{Recalculate inter-cluster weights $w(c_{new}, c_i)$ for all $c_i \in cluster\_map \setminus \{c_{new}\}$}
            \ENDIF
        \UNTIL{all inter-cluster weights $w(c_i, c_j) \leq \tau$}
    \STATE{\textit{Return} : $cluster\_map$}
    \end{algorithmic}
\end{algorithm}

Algorithm \ref{algo:comm_col_cluster_algo} presents our proposed communication and colocation-based clustering approach. It obtains the centralized topology graph \(\mathcal{G'} = (\mathcal{V}, \mathcal{L_{\mathrm{comm}}})\) consisting of the set of microservices \( M \) and the set of edge devices \( E \) as vertices, and the set of communication edges between microservices \( L_{comm} \), together with function $\delta(m)$ which maps each microservice $m$ to its deployment node and $\tau$ which is the threshold to merge clusters as inputs. As an edge device is the smallest unit that groups colocated microservices, the algorithm starts by initializing $cluster\_map$, which maintains the cluster to edge device mapping, such that there is a cluster corresponding to microservices deployed in each edge device. However, since microservices are placed primarily to satisfy QoS requirements, rather than based on communication frequency, there would be communication dependencies between initial clusters. Therefore, our algorithm next attempts to group such devices with high communication frequencies. As detailed in step \ref{comm_col_cluster:weight_calc} of Algorithm \ref{algo:comm_col_cluster_algo}, we calculate the inter-cluster communication weight for each pair of clusters. This weight represents the communication frequency between the microservices deployed in those clusters. Subsequently, the algorithm performs an iterative merging procedure. At each step, it identifies the pair of clusters with the highest inter-cluster communication weight. If this weight exceeds the threshold $\tau$, the two clusters are merged into a single cluster, and the $cluster\_map$ is updated accordingly. The inter-cluster communication weights involving the newly formed cluster are then recalculated. This process continues until no inter-cluster weight exceeds the threshold $\tau$, ensuring that clusters are only merged when strong communication relationships exist. Finally, the algorithm outputs the updated $cluster\_map$, which contains the final cluster-to-edge device mapping.

\begin{algorithm}[t]
\setstretch{0.9}
    \caption{Topology subgraph formation for decentralized RCL modules} 
    \label{algo:sub_topology_graph}
    \begin{algorithmic}[1]
        \STATE{\textit{Input}: Final $cluster\_map$ from Algorithm~\ref{algo:comm_col_cluster_algo}}
        \STATE{\textit{Input}: Centralized topology graph $G' = (V, L_{comm})$}
        \STATE{\textit{Output}: Set of topology subgraphs $\{ G'_i \}$, one for each cluster $c_i$}
        \FOR{each cluster $c_i$ in $cluster\_map$}
            \STATE{$V_{c_i} \gets \{ m \in V \mid \delta(m) \in c_i \}$} \COMMENT{Vertices: microservices in cluster $c_i$}
            \STATE{$L_{c_i} \gets \{ (m_p, m_q) \in L_{comm} \mid m_p \in V_{c_i} \ \land \ m_q \in V_{c_i} \}$} \COMMENT{Intra-cluster edges}
            \STATE{$proxy\_nodes \gets \emptyset$}
            \FOR{each $(m_p, m_q) \in L_{comm}$ where $m_p \in V_{c_i}$ and $m_q \notin V_{c_i}$}
                \STATE{Create a new shadow node $s_{m_q}$ representing $m_q$ inside cluster $c_i$}
                \STATE{$proxy\_nodes \gets proxy\_nodes \cup \{ s_{m_q} \}$}
                \STATE{$L_{c_i} \gets L_{c_i} \cup \{ (m_p, s_{m_q}) \}$} \COMMENT{Redirect outgoing edge to proxy node}
            \ENDFOR
            \STATE{$V_{c_i} \gets V_{c_i} \cup proxy\_nodes$}
            \STATE{Form topology subgraph $G'_i = (V_{c_i}, L_{c_i})$}
        \ENDFOR
        \STATE{\textit{Return}: Set of topology subgraphs $\{ G'_i \}$, one for each cluster $c_i$}
    \end{algorithmic}
\end{algorithm}

Once the final set of clusters is obtained, we construct the topology subgraph for the decentralized RCL module corresponding to each cluster. This process is explained in Algorithm \ref{algo:sub_topology_graph}. It obtains the $cluster\_map$ which is the output of Algorithm \ref{algo:comm_col_cluster_algo} together with the centralized topology graph \(\mathcal{G'} = (\mathcal{V}, \mathcal{L_{\mathrm{comm}}})\) consisting of the set of microservices \( M \) and the set of edge devices \( E \) as vertices, and the set of communication edges between microservices \( L_{comm} \), as inputs. For a given cluster \(c_i\), the vertices of its topology subgraph, \(G'_i\), consist of all microservices deployed on the edge devices assigned to \(c_i\). The intra-cluster edges are defined as the communication edges between these microservices in the centralized topology graph, \(G'\). In addition, to preserve the connectivity information with external microservices while ensuring that the sub-topology remains self-contained, each communication edge originating from a microservice in \(c_i\) to a microservice outside \(c_i\) is redirected to a shadow node placed within \(G'_i\). This shadow node symbolically represents the remote endpoint and serves as the destination for the redirected outgoing edge. The resulting topology subgraph therefore captures both the internal structure of the cluster and its interaction points with the rest of the system, enabling the decentralized RCL module to operate independently while still considering external dependencies.

\begin{figure}[t]
\centerline{\includegraphics[width=\columnwidth]{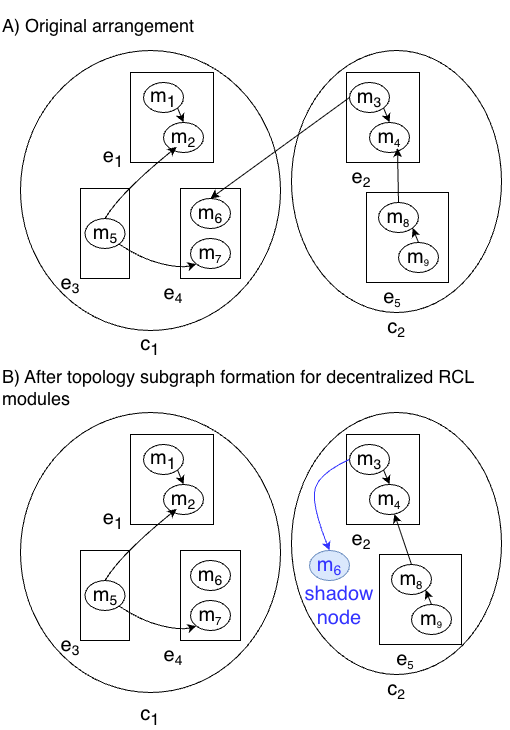}}
\caption{Inter-cluster communication edge representation using a shadow node}
\label{subtopology_graph_formation}
\end{figure}

To illustrate how an inter-cluster communication edge is represented using a shadow node in the source cluster, we refer to Figure~\ref{subtopology_graph_formation}. In this example, clusters \(c_1\) and \(c_2\) are connected by an inter-cluster communication edge. The topology subgraph \(G'_1\) for cluster \(c_1\) contains as vertices all microservices deployed within \(c_1\), along with the communication edges between them. Similarly, the topology subgraph \(G'_2\) for cluster \(c_2\) contains all microservices deployed within \(c_2\) and their internal communication edges. In addition, \(G'_2\) includes a shadow node representing the destination microservice in \(c_1\) that is the endpoint of the outgoing inter-cluster link from \(c_2\). The outgoing edge from the originating microservice in \(c_2\) is redirected to this shadow node, allowing \(G'_2\) to capture the external communication dependency without directly incorporating microservices from outside its own cluster.

The above-mentioned inter-cluster communications are expected to be infrequent, and the occurrence of anomalies along such inter-cluster edges is even rarer. Each cluster is deliberately constructed to capture the majority of communication and colocation relationships, thereby ensuring that most anomalies propagate entirely within a single cluster. Consequently, RCL can typically be performed within the cluster in which the anomaly is propagated. Upon anomaly detection, the corresponding topology subgraph \(G'_i\) is populated with the anomalous metrics associated with its microservices. Subsequently, the personalisation vector \(S_i\) and the transition probability matrix \(P_i\) are constructed using the novel anomaly scoring mechanism introduced in section \ref{subsec:anomaly_scoring_edge}.

\subsection{Decentralized execution of the Personalized PageRank algorithm}
\label{subsec:decentralised_ppr}

There are two possible scenarios when performing PPR, depending on whether the incoming edges to shadow nodes are detected as anomalous. In most cases, due to communication and colocation-based clustering, anomaly propagation remains confined within a single cluster. In such cases, PPR can be executed locally within the cluster according to the formulation in subsection \ref{subsec:ppr}, iterating until convergence and returning the microservice with the highest probability in \(r\). In this scenario, both \(P_i\) and \(S_i\) are constructed using only the original (non-shadow) nodes in the cluster.

Although the clustering strategy is designed to minimise inter-cluster anomaly propagation, rare cases may still occur where anomalies traverse less frequent communication paths spanning multiple clusters. These cases are detected when incoming edges to shadow nodes are identified as anomalous. When this happens, the anomaly propagation path between clusters is considered disconnected, and the inter-cluster peer-to-peer (p2p) approximation process is triggered.

\begin{figure}[t]
\centerline{\includegraphics[width=\columnwidth]{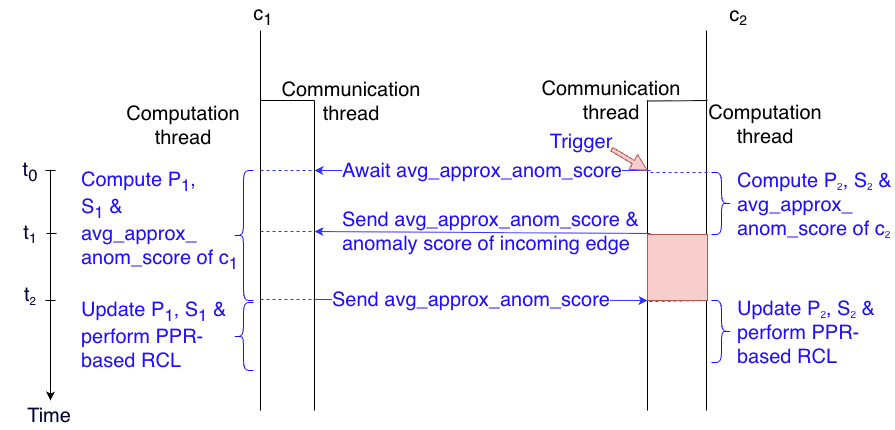}}
\caption{Inter-cluster peer-to-peer (p2p) approximation process}
\label{p2p_approx_interaction_diagram}
\end{figure}

Consider the example in Figure \ref{subtopology_graph_formation} together with the corresponding interaction diagram in Figure \ref{p2p_approx_interaction_diagram}. At time \(t_0\), the computation threads of both clusters \(c_1\) and \(c_2\) independently begin computing their respective \(P_i\) and \(S_i\). If an anomaly is detected on the incoming edge to the shadow node in \(c_2\), that indicates a disconnection in the anomaly propagation path and triggers the inter-cluster p2p approximation process. At that time (i.e., \(t_0+\delta_1\)), the communication thread of cluster \(c_2\) sends a \(wait\_message\) to \(c_1\), indicating its intent to provide an average approximate anomaly score. This approximation is used to reduce communication overhead while still capturing the anomaly influence of \(c_2\) across iterations. While the communication thread is sending the \(wait\_message\), the RCL module (i.e., the computation thread) in \(c_2\) continues to compute the anomaly scores required to form \(P_2\) and \(S_2\) and uses them to calculate its average approximate anomaly score as: 

\begin{equation}
\label{eq:avg_approx_anom_score}
\mathrm{avg\_approx\_anom\_score} = \frac{\sum\limits_{x \in S} x}{|S|}
\end{equation}

After completing the computation (i.e., at time \(t_1\)), this value is sent to \(c_1\) along with the anomaly score of the incoming edge to the shadow node. 

Upon receiving the \(wait\_message\) from \(c_2\) at time \(t_0+\delta_2\), \(c_1\) continues to compute \(P_1\) and \(S_1\), while awaiting \(c_2\)’s average approximate anomaly score. It also calculates its own average approximate anomaly score in the same way as equation \ref{eq:avg_approx_anom_score} and sends it to \(c_2\) when ready at time \(t_2\). Note that one cluster must wait for the other to complete this computation, since \(t_1 \neq t_2\). In the depicted example, \(t_2 > t_1\), and thus \(c_2\) waits, as indicated by the red block between \(t_1\) and \(t_2\).

Once both clusters have exchanged their average approximate anomaly scores, they update their \(P_i\) and \(S_i\) (a straightforward update to the corresponding matrix and vector) by incorporating the other cluster as a node. Both clusters then run their PPR-based RCL algorithms independently, as in the standard scenario. This procedure is a modified version of the JXP algorithm \cite{parreria2006jxp}, an efﬁcient and decentralized PageRank approximation used in peer-to-peer web search networks. Without loss of generality, this communication protocol is extendable to scenarios with more than two clusters.

Finally, during decision making, one cluster \(c_i\) (e.g., \(c_1\) in our case) typically identifies a microservice within its boundary as the root cause, while other clusters (e.g., \(c_2\)) point to \(c_i\) as the origin of the anomaly. In this case, the microservice identified by \(c_i\) is selected as the final root cause. While this is the majority case, in practical scenarios (expected to be extremely rare) where multiple clusters identify distinct local microservices as the most likely causes, an inter-cluster result aggregation mechanism (explained next) is applied to reconcile these decisions.  

After completing their independent PPR executions, each cluster \(c_i\) produces a ranking vector
\[
r_i = [r^{m_1}_i, r^{m_2}_i, \ldots, r^{m_k}_i, r^{c_1}_i, r^{c_2}_i, \ldots, r^{c_n}_i]
\]
where the first \(k\) elements correspond to its internal microservices (\(m_j\)) and the remaining \(n\) elements represent connected clusters (\(c_j\)). The term \(r^{c_j}_i\) denotes the anomaly likelihood assigned to cluster \(c_j\) by cluster \(c_i\).

Next, each cluster integrates its internal anomaly profile and its inferred external influences from other connected clusters to construct its global perspective vector as follows:
\[
r^{\text{global}}_{i} = [r^{\text{local}}_{i}, r^{c_1\text{ influence}}_{i}, r^{c_2\text{ influence}}_{i}, \ldots, r^{c_n\text{ influence}}_{i}]
\]

Here, \(r^{\text{local}}_i = [r^{m_1}_i, r^{m_2}_i, \ldots, r^{m_k}_i]\), which corresponds to the internal anomaly profile, represents the anomaly likelihoods of microservices within \(c_i\). For each connected cluster \(c_j\), the influence component \(r^{c_j\text{ influence}}_{i}\) is obtained as:
\[
r^{c_j\text{ influence}}_{i} = r^{c_j}_i \cdot r^{\text{local}}_{j}
\]
The idea behind this formulation is to distribute the anomaly likelihood assigned to cluster \(c_j\) (by cluster \(c_i\)) proportionally across \(c_j\)’s local anomaly scores, capturing how \(c_i\) perceives \(c_j\)’s influence.

Finally, the global perspectives from all clusters are averaged to derive a unified anomaly probability vector:
\[
r_i^{\text{combined}} = \frac{1}{N} \sum_{i=1}^N r^{\text{global}}_{i}
\]
The microservice with the highest likelihood in \(r_i^{\text{combined}}\) is selected as the final root cause. This aggregation ensures that all participating clusters contribute to the final decision, enhancing robustness in rare cases of inter-cluster anomaly propagation.


\section{Performance Evaluation}
\label{sec:perf_eval}

In this section, we discuss the evaluation results of the proposed decentralized RCL approach. First, in subsection \ref{subsec:exp_setup}, we explain the details of the experimental setup used for evaluation together with implementation details. Next, in subsection \ref{subsec:main_decentral_eval}, we discuss the results of evaluating the proposed method against its centralized counterpart, which is commonly referenced in the existing literature \cite{wu2020microrca,tian2023microgbpm} and serves as our baseline approach. Following this, in subsection \ref{subsec:further_analysis_eval}, we conduct a further analysis of our decentralized results in the context of both single-cluster and multi-cluster anomaly propagation cases. Finally, subsection~\ref{subsec:eval_discussion} presents a broader discussion of the findings, highlighting the implications of decentralization in terms of localization accuracy, efficiency, and communication overhead within edge environments.

\subsection{Experimental setup and implementation details}
\label{subsec:exp_setup}

We evaluated the proposed decentralized RCL approach using the publicly available MicroCERCL dataset\footnote{https://github.com/WDCloudEdge/MicroCERCL} \cite{zhu2024microcercl}. This dataset represents the first large-scale benchmark for cloud–edge collaborative microservice systems and remains the most comprehensive hybrid deployment dataset available to date. It contains data collected from 81 microservices belonging to four applications: SockShop, Hipster, Bookinfo, and the newly introduced AI-Edge. These microservices are deployed across four cloud servers and two groups of two edge servers, following a communication frequency-based application placement strategy.

To reflect realistic production environments, the dataset integrates a wide range of anomalies. Using ChaosMesh, the authors injected application-level anomalies, such as CPU resource exhaustion in containers, memory leaks, and network latency. Additionally, Linux kernel traffic control (TC) was employed to simulate kernel-level network failures—including packet loss, duplication, corruption, disorder, delay, and jitter—between cloud and edge nodes. Consequently, the dataset captures both communication- and colocation-induced anomaly propagation patterns.

From the available dataset, we selected 383 scenarios that provide adequate diversity in fault types. Among them, 237 scenarios (denoted SH) correspond to cases where the root-cause microservice belongs to the SockShop application, while the remaining 146 scenarios (denoted HH) involve root causes from the Hipster application. For each scenario, we extracted trace information to construct topology graphs, and corresponding metrics were used for anomaly detection and RCL. 

All experiments, including hyperparameter tuning, were conducted on the Spartan HPC cluster\footnote{https://dashboard.hpc.unimelb.edu.au/}. The implementation was performed in Python 3.10 using PyTorch 2.2\footnote{https://pytorch.org/}, SciPy 1.13\footnote{https://scipy.org/}, and Scikit-learn 1.1\footnote{https://scikit-learn.org/}. 

To construct decentralized subgraphs, we first applied Algorithm \ref{algo:comm_col_cluster_algo} to cluster the edge devices in each scenario based on their communication and colocation dependencies. The resulting cluster assignments were then used to generate corresponding topology subgraphs using Algorithm \ref{algo:sub_topology_graph}. Both algorithms were implemented in Python. To determine the optimal threshold for merging clusters (denoted as $\tau$), which is a key hyperparameter required for Algorithm \ref{algo:comm_col_cluster_algo}, we employed Tree-structured Parzen Estimator (TPE)-based Bayesian optimization \cite{bergstra2011tpe}. Across all scenarios, this process produced between 2 and 4 clusters per scenario.

For anomaly detection, we adopted the BIRCH clustering-based algorithm from Scikit-learn, following the configuration recommended by the MicroCERCL authors. Specifically, the anomaly sensitivity threshold \(\beta\) was set to 0.07, balancing anomaly detection accuracy and noise reduction.

The proposed anomaly scoring mechanism (Section \ref{subsec:anomaly_scoring_edge}) was implemented using the SciPy library. Subsequently, the PPR-based RCL algorithm—used in both decentralized and centralized variants—was implemented in Python. Its hyperparameters, \( \alpha \) and \( \varepsilon \), were tuned using the same TPE-based Bayesian optimization approach \cite{bergstra2011tpe}.

The next subsection presents a detailed comparison between our decentralized RCL approach and its centralized baseline.

\subsection{Evaluation against the centralized baseline}
\label{subsec:main_decentral_eval}

Before evaluating the decentralized RCL approach, we first validate the centralized variant of our method, which utilizes the PPR-based localization algorithm. This step ensures that our implementation achieves comparable performance to the existing MicroCERCL benchmark \cite{zhu2024microcercl}, thereby establishing it as a reliable baseline for subsequent comparison.

\begin{table*}[h] 
\centering 
\footnotesize 
\caption{Comparison of centralized RCL results on the MicroCERCL dataset} 
\label{tab:microcercl_vs_ours} 
\begin{tabular}{lccccccc} 
\toprule 
\textbf{Application} & \textbf{Acc@1} & \textbf{Acc@2} & \textbf{Acc@3} & \textbf{Acc@5} & \textbf{Acc@10} & \textbf{MAR} & \textbf{MRR} \\ \midrule \multicolumn{8}{l}{\textit{MicroCERCL (GNN-based centralized)}} \\ HH & 0.632 & 0.756 & 0.796 & 0.833 & 0.896 & – & – \\ SH & 0.607 & 0.732 & 0.792 & 0.849 & 0.907 & – & – \\ \midrule \multicolumn{8}{l}{\textit{Our baseline (PPR-based centralized)}} \\ HH & 0.4315 & 0.6507 & 0.8082 & 0.9726 & 1.0000 & 2.2877 & 0.6352 \\ SH & 0.4979 & 0.7300 & 0.8270 & 0.9451 & 0.9958 & 2.1688 & 0.6817 \\ 
\bottomrule 
\end{tabular} 
\end{table*}

Table~\ref{tab:microcercl_vs_ours} compares the results of our centralized PPR-based approach with those reported in the MicroCERCL paper, where a GNN-based centralized RCL model was used. The performance is evaluated using the standard Accuracy@k metric, which measures the proportion of test cases where the true root-cause microservice appears within the top-$k$ ranked predictions. In our evaluation, we report results for $k = 1, 2, 3, 5,$ and $10$, covering both precise and broader localization ranges. In addition, we report the Mean Average Rank (MAR), which indicates the average position of the true root cause (lower values are better), and the Mean Reciprocal Rank (MRR), which reflects how early the correct root cause appears in the ranked list (higher values are better).

As shown in Table~\ref{tab:microcercl_vs_ours}, our centralized PPR-based approach demonstrates competitive localization performance relative to MicroCERCL's GNN-based model. While the top-1 accuracy (Accuracy@1) is moderately lower, the performance steadily improves at larger values of $k$, reaching near-perfect Accuracy@10 in both application cases. This trend indicates that PPR effectively ranks the true root-cause microservice within a small set of top candidates, even without model training or feature learning.

The average ranking performance, reflected by MAR and MRR, further confirms this observation. For both applications, the MAR values of 2.29 (HH) and 2.17 (SH) indicate that the true root cause typically appears near the second position in the ranked list. Meanwhile, MRR values of 0.64 and 0.68 imply that in many scenarios the correct root cause appears at rank 1 or 2, thus remaining consistently close to the top of the ranking list. Overall, these results establish our centralized PPR-based implementation as a strong and reliable baseline for comparison with the proposed decentralized RCL approach.

Having established the centralized PPR-based RCL model as a valid baseline, we next evaluate the performance of our proposed decentralized RCL approach. This evaluation focuses on two key aspects: localization accuracy and localization efficiency. Similar to the centralized evaluation, we use the Accuracy@k (for $k = 1, 2, 3, 5,$ and $10$), Mean Average Rank (MAR), and Mean Reciprocal Rank (MRR) metrics to measure localization accuracy. In addition, to assess the efficiency improvement achieved through decentralization, we introduce two new metrics: Average Time Reduction Percentage and Total Time Reduction Percentage.

The Average Time Reduction Percentage quantifies the mean percentage reduction in localization time across all scenarios. It is defined as
\[ 
\frac{1}{N} \sum_{i=1}^{N} 
\left( 
\frac{t_{\text{centralized}}^{(i)} - t_{\text{decentralized}}^{(i)}}{t_{\text{centralized}}^{(i)}}
\times 100
\right),
\]
where $t_{\text{centralized}}^{(i)}$ and $t_{\text{decentralized}}^{(i)}$ denote the localization times under the centralized and decentralized settings for the $i^{th}$ scenario, respectively, and $N$ is the total number of scenarios. This metric reflects the average per-scenario improvement achieved through decentralization.

The Total Time Reduction Percentage, on the other hand, considers the aggregate time across all scenarios, defined as
\[
\frac{\sum t_{\text{centralized}} - \sum t_{\text{decentralized}}}{\sum t_{\text{centralized}}} \times 100.
\label{eq:macro_time_reduction}
\]

Together, these two metrics capture both per-scenario consistency (Average Time Reduction Percentage) and overall efficiency gain (Total Time Reduction Percentage).

\begin{table*}[h]
\centering
\footnotesize
\caption{Centralized vs. decentralized RCL performance}
\label{tab:centralized_vs_decentralized}
\begin{tabular}{lccccccccc}
\toprule
\textbf{Application} & \textbf{Acc@1} & \textbf{Acc@2} & \textbf{Acc@3} & \textbf{Acc@5} & \textbf{Acc@10} & \textbf{MAR} & \textbf{MRR} & 
\begin{tabular}[c]{@{}c@{}}\textbf{Average Time}\\\textbf{Reduction (\%)}\end{tabular} &
\begin{tabular}[c]{@{}c@{}}\textbf{Total Time}\\\textbf{Reduction (\%)}\end{tabular} \\
\midrule
\multicolumn{10}{l}{\textit{Centralized baseline}} \\
HH & 0.4315 & 0.6507 & 0.8082 & 0.9726 & 1.0000 & 2.2877 & 0.6352 & -- & --\\
SH & 0.4979 & 0.7300 & 0.8270 & 0.9451 & 0.9958 & 2.1688 & 0.6817 & -- & --\\
\midrule
\multicolumn{10}{l}{\textit{Proposed decentralized approach}} \\
HH & 0.5616 & 0.7671 & 0.8767 & 0.9863 & 1.0000 & 1.8630 & 0.7295 & 17.41 & 21.34 \\
SH & 0.6076 & 0.8650 & 0.9325 & 0.9831 & 1.0000 & 1.6582 & 0.7733 & 33.22 & 33.59 \\
\bottomrule
\end{tabular}
\end{table*}

As shown in Table~\ref{tab:centralized_vs_decentralized}, the decentralized approach consistently outperforms the centralized baseline across both application groups (HH and SH). In terms of accuracy, the decentralized method improves Accuracy@1 from 0.43 (HH) and 0.50 (SH) to 0.56 and 0.61, respectively, while achieving even higher scores for larger $k$ values. Both MAR and MRR values also demonstrate notable gains, indicating that the true root-cause microservice appears higher in the ranked list under decentralized processing. This improvement in accuracy can be attributed to the noise reduction effect introduced by decentralization: by performing RCL within smaller, communication-aware clusters, each subgraph captures more localized dependencies and avoids the propagation of irrelevant information from distant microservices.

In terms of efficiency, the results indicate substantial time reductions, with 17–21\% improvement for HH and over 33\% for SH in both average and total time reduction percentages. These results highlight that decentralization not only maintains but, in many cases, enhances localization accuracy, while significantly reducing localization time, thus achieving a more effective balance between accuracy and efficiency.

These results collectively confirm that decentralization leads to clear gains in both accuracy and efficiency compared to the centralized baseline. However, the extent of these improvements can depend on how anomalies propagate within the system. To better understand the influence of anomaly propagation characteristics, the next subsection analyzes the decentralized results in greater depth by separating the evaluation into single-cluster and multi-cluster cases.

\subsection{Detailed analysis of decentralized results}
\label{subsec:further_analysis_eval}

While the previous subsection presented the overall performance of the decentralized RCL approach, a more granular analysis could provide deeper insights into its behavior under different anomaly propagation patterns. Specifically, we categorize the 383 evaluation scenarios (237 SH and 146 HH) into single-cluster propagation and multi-cluster propagation cases. This categorization enables us to assess whether the decentralized approach consistently maintains localization accuracy and efficiency regardless of whether the anomaly remains confined to a single cluster or propagates across multiple clusters.

\begin{table*}[h]
\centering
\footnotesize
\caption{Decentralized RCL performance broken down by anomaly propagation type}
\label{tab:decentralized_breakdown}
\setlength{\tabcolsep}{4pt} 
\renewcommand{\arraystretch}{1.2} 
\begin{tabular}{llcccccccccc}
\toprule
\textbf{Propagation Type} & \textbf{Case} & \textbf{\# Scenarios} &
\textbf{Acc@1} & \textbf{Acc@2} & \textbf{Acc@3} & \textbf{Acc@5} & \textbf{Acc@10} &
\textbf{MAR} & \textbf{MRR} &
\makecell{\textbf{Average Time}\\\textbf{Reduction (\%)}} &
\makecell{\textbf{Total Time}\\\textbf{Reduction (\%)}} \\
\midrule
Single-cluster & HH & 105 & 0.5333 & 0.7429 & 0.8476 & 0.9810 & 1.0000 & 1.9714 & 0.7079 & 23.12 & 24.02 \\
               & SH & 206 & 0.5680 & 0.8447 & 0.9223 & 0.9806 & 1.0000 & 1.7379 & 0.7488 & 33.89 & 34.44 \\
Multi-cluster  & HH & 41  & 0.6341 & 0.8293 & 0.9512 & 1.0000 & 1.0000 & 1.5854 & 0.7846 & 2.77  & 14.07 \\
               & SH & 31  & 0.8710 & 1.0000 & 1.0000 & 1.0000 & 1.0000 & 1.1290 & 0.9355 & 28.79 & 27.41 \\
\bottomrule
\end{tabular}
\end{table*}

Table \ref{tab:decentralized_breakdown} summarizes the results for both categories, reporting the same set of accuracy and efficiency metrics — Accuracy@k (for $k = 1, 2, 3, 5,$ and $10$), Mean Average Rank (MAR), Mean Reciprocal Rank (MRR), and the average/total time reduction percentages.

The majority of scenarios (311 out of 383) fall under the single-cluster anomaly propagation category, where the anomaly originating from the root-cause microservice propagates only within its local cluster. In these cases, our decentralized method demonstrates a substantial efficiency improvement — achieving over 23–34 \% reduction in localization time on average — while maintaining strong accuracy levels (Accuracy@3 \textgreater 0.84 for both HH and SH). These results demonstrate that when the anomaly impact is localized, decentralized inference efficiently uses intra-cluster communication to identify the root cause with minimal overhead.

For the remaining multi-cluster propagation cases (72 out of 383), where anomalies propagate across clusters and the algorithm engages in inter-cluster peer-to-peer (P2P) approximation, the decentralized approach continues to perform competitively. Accuracy metrics remain high (Accuracy@1 $\geq$ 0.63 for HH and $\geq$ 0.87 for SH), validating that the cross-cluster coordination process is effective. Among the 72 multi-cluster propagation cases, a very small subset required the additional inter-cluster result aggregation step—specifically when both participating clusters independently identified local anomalies as potential root causes. In these cases, the aggregation step ensured consistent decision-making without sacrificing accuracy. The resulting accuracy values remained stable (Accuracy@1 $\geq$ 0.63 for HH and $\geq$ 0.87 for SH), indicating that the inter-cluster result aggregation mechanism effectively reconciles results from different clusters.

As expected, the efficiency gains are somewhat lower (2.77–28.79\%) in cross-cluster scenarios due to the additional latency introduced by inter-cluster communication. The impact of the inter-cluster result aggregation mechanism, however, is minimal: since it requires only a single exchange of low-dimensional ranking vectors, its communication overhead is negligible compared to the total P2P coordination time. The slight reduction in overall efficiency is therefore primarily attributed to inter-cluster message synchronization rather than to aggregation itself. Despite this, the method still achieves a notable reduction in localization time while maintaining, or even improving, localization accuracy compared to the centralized baseline.

Overall, this breakdown illustrates that the proposed decentralized approach adapts well to both local and distributed anomaly propagation scenarios, sustaining accuracy while offering consistent reductions in localization time.

\subsection{Discussion}
\label{subsec:eval_discussion} 

In summary, the evaluation results demonstrate that the proposed decentralized RCL approach achieves comparable or higher localization accuracy than the centralized baseline, while significantly reducing localization time. The approach performs consistently well across both single-cluster and multi-cluster anomaly propagation scenarios, confirming its effectiveness under diverse anomaly propagation conditions. 

The results also validate that the inter-cluster peer-to-peer approximation process—designed for multi-cluster anomaly propagation—introduces minimal communication overhead and exerts negligible impact on localization time. This process requires only a one-time exchange of average approximate anomaly scores between connected clusters, involving a limited number of lightweight message exchanges overall (\(wait\_message\) and \(avg\_approx\_anom\_score\)). Empirical evidence confirms that these exchanges incur almost no waiting periods, as most computations proceed in parallel across clusters, ensuring efficient utilization of resources during coordination.

Moreover, in the rare cases where multiple clusters identify distinct local anomalies, the results show that the inter-cluster result aggregation mechanism effectively reconciles these outcomes with minimal overhead. Since this step operates on compact ranking vectors rather than raw monitoring data, both communication and computation costs remain negligible. This confirms that the efficiency benefits of decentralization are preserved even in cross-cluster anomaly propagation scenarios.

Finally, beyond the observed improvements in localization time due to faster convergence of the PPR-based RCL approach, the decentralized design is expected to further reduce latency in real deployments by executing RCL at the edge device level. This minimizes data transfer delays and lowers communication overhead compared to traditional centralized approaches, where large volumes of monitoring data must be transported to a central node for analysis. 

\section{Conclusions and Future Work}
\label{sec:conclusions}

This paper proposed a decentralized root cause localization (RCL) approach for microservice-based IoT applications deployed in edge computing environments. By leveraging the Personalized PageRank (PPR) algorithm within communication and colocation-aware clusters, our approach reduces localization time without compromising accuracy. The results demonstrate that performing RCL closer to edge devices significantly improves efficiency, achieving comparable or better accuracy than centralized baselines while reducing communication overhead. The inter-cluster peer-to-peer approximation process further ensures robustness in rare cases of multi-cluster anomaly propagation, confirming the scalability of the proposed decentralized design.

While effective under communication-aware placement strategies—where microservices that frequently communicate are co-located—the proposed approach currently assumes that such placement is at least used as a secondary objective alongside QoS-aware placement. This is because the current design does not generalize to scenarios involving frequent bi-directional inter-cluster communications. Ideally, such highly interactive microservices should belong to the same cluster. Therefore, an important direction for future work is to extend the proposed approach to accommodate other placement techniques, such as purely QoS-driven or adaptive placement strategies, while maintaining efficiency and accuracy. We also plan to explore optimization opportunities in subsequent stages of the anomaly management pipeline, particularly root cause analysis (RCA) and joint RCL–RCA techniques, with a focus on enhancing efficiency and alignment with edge-specific constraints.


\bibliographystyle{IEEEtran}
	
\bibliography{reference}

\end{document}